\documentclass[12pt,preprint]{aastex}
\begin{document}
\title{High Jet Efficiency and Simulations of Black Hole Magnetospheres}
\author{Brian Punsly\altaffilmark{1}} \altaffiltext{1}{4014 Emerald Street No.116, Torrance CA,
USA 90503 and ICRANet, Piazza della Repubblica 10 Pescara 65100,
Italy, brian.punsly@verizon.net or brian.punsly@comdev-usa.com}

\begin{abstract}
This article reports on a growing body of observational evidence
that many powerful lobe dominated (FR II) radio sources likely have
jets with high efficiency. This study extends the maximum efficiency
line (jet power $\approx$ 25 times the thermal luminosity) defined
in Fernandes et (2010) so as to span four decades of jet power. The
fact that this line extends over the full span of FR II radio power
is a strong indication that this is a fundamental property of jet
production that is independent of accretion power. This is a
valuable constraint for theorists. For example, the currently
popular "no net flux" numerical models of black hole accretion
produce jets that are 2 to 3 orders of magnitude too weak to be
consistent with sources near maximum efficiency.
\end{abstract}


\section{Introduction}Relativistic jets emanating from AGN (active galactic nuclei) exist in a variety of
strengths. Most quasars are radio quiet objects that have either no
measurable jets or jets that are so weak that they often cannot
propagate out of the host galaxy. About $\gtrsim 10\%$ of AGN
have highly luminous radio jets of which $\approx 20\%$ are classic
FR II radio sources defined by jets that propagate hundreds of kpc,
terminating in lobes of plasma with similar linear extent
\citep{dev06}. The energy flux in these jets, $Q$, can be enormous
with many independent estimates finding long term time averages,
$\overline{Q}\gtrsim 10^{47}\mathrm{ergs/sec}$ \citep{wil99,pun10}.
These jets are not perfectly steady, so there are episodes in which
the instantaneous power, $Q(t)$, must be even larger. In this
Letter, an attempt is made to expand on and consolidate the evidence
for large jet power that has been accumulating in the literature
since 2006.
\par The second section quantifies $Q$ relative to the dynamics of the accreting gas.
From observations, one can estimate the ratio of $Q$ to the thermal
luminosity of the accretion flow, $L_{bol}$, defined as $R \equiv
Q/L_{bol}$ or with respect to the mass accretion rate $Q \equiv
\eta_{Q} \dot{M} c^{2}$, where $\eta$ is called the efficiency.
Strong radio sources can be described in terms of a concept of
maximum jet efficiency that was introduced by \citet{fer10}, i.e a
maximum value of $R$ or $\eta_{Q}$ for radio jets. The maximum jet
efficiencies implied by various lines of research are compiled. The
results are analyzed for consistency and are critically examined in
terms of the assumptions and limitations of each method.
\par In the third section, we discuss the high efficiency sources from
section 2 in the context of current numerical work, the 3-D
numerical simulations of MHD (magnetohydrodynamic) accretion flows
around black holes with no net magnetic flux. The nexus between
these simulations and observation is that the simulated $Q$ is
always expressible in units of $\dot{M} c^{2}$ by all research
groups. By making reasonable assumptions about $\eta_{th}$, the
thermal efficiency ($L_{bol} = \eta_{th}\dot{M} c^{2}$), as derived
by accretion disk theory and new turbulent MHD simulations, one can
compare the observations with the simulations.
\section{Evidence for High Jet Efficiency}Ever since the seminal work
of \citet{raw91}, astrophysicists have been trying to estimate the
enormous energy flux that feeds the radio lobes in FR II radio
sources and relate it to the thermal luminosity of the accretion
flow. The three most viable options for estimating $Q$ are either
based on the low frequency (151 MHz) flux from the radio lobes on
100 kpc scales (eg. \citet{wil99}), or the work done creating the
cavities that are carved out of the intra-cluster medium by the
expanding radio lobes (eg. \citet{bir04,mcn10}), or models of the
broadband Doppler boosted synchrotron and inverse Compton radiation
spectra associated with the relativistic parsec scale jet (eg.
\citet{ghi10}). Each method has its advantage. The 151 MHz method is
the most widely applicable, all that is needed is a radio spectrum.
A disadvantage is that it involves long term time averages,
$\overline{Q}$, that do not necessarily reflect the current state of
quasar activity. The second method is also not contemporaneous, yet
more accurate in principle than the first, but is restricted to low
redshift sources with deep X-ray observations. The last method is
contemporaneous, so one can define a directly interpretable ratio of
jet to accretion thermal power, $R(t)=Q(t)/L_{bol}$. However, one is
forced to deal with estimating the large Doppler enhancement factor,
$\delta $, which is a potential source of large uncertainty (the jet
luminosity scales like $\delta^{4}$). In this section, using these
three methods, we expand on the notion of a maximum jet efficiency
defined in \citet{fer10} that is based on 151 MHz flux estimate.
   \par In Figure 1, the black squares are a scatter plot,
$\overline{Q}$ versus $L_{bol}$, of the complete sample of FRII
narrow line radio galaxies (NLRGs) from \citet{fer10}. The low
frequency selected sample is limited to $0.9<z<1.1$, as a compromise
between having sufficient cosmic volume to find strong radio sources
and being sufficiently close so that these sources can be detected
in the IR. $L_{bol} = 8.5 \nu L_{\nu}(12\mu \mathrm{m})$ is computed
from the IR luminosity at 12 $\mu \mathrm{m}$ \citet{ric06}. They
define a diagonal line at $\overline{R}\equiv \overline{Q}/L_{bol}
=25$ (the black solid line in Figure 1), the maximum efficiency
line. The dashed blue vertical line represents the approximate
dividing line between Seyfert 1 galaxy and quasar luminosity ($M_{V}
= -23$). The dashed vertical orange line represents the dividing
line between solely Seyfert 1 galaxies and a mixture of LLAGNs (low
luminosity AGN) and weaker Seyfert 1 galaxies \citep{lho05}. In
contrast to Seyfert 1 galaxies and quasars, the LLAGNs do not have a
strong "blue bump" in their spectra, the signature of strong thermal
emission from an accretion flow \citep{lho05,sun89}. The LLAGNs are
estimated to have inefficient modes of accretion, expressed in terms
of the Eddington luminosity as $L_{bol}/L_{Edd} < 10^{-5}$ in
contrast to quasars and Seyfert galaxies which typically have
$10^{-2} < L_{bol}/L_{Edd} < 1$ \citep{lho05,sun89}. Thus, Figure 1
indicates that the NLRGs in the \citet{fer10} sample are likely to
have a central black hole in a high efficiency accretion state
typical of a quasar or Seyfert 1 galaxy, but the optical/UV core is
hidden.
\begin{figure}
\includegraphics[width=160 mm, angle= 0]{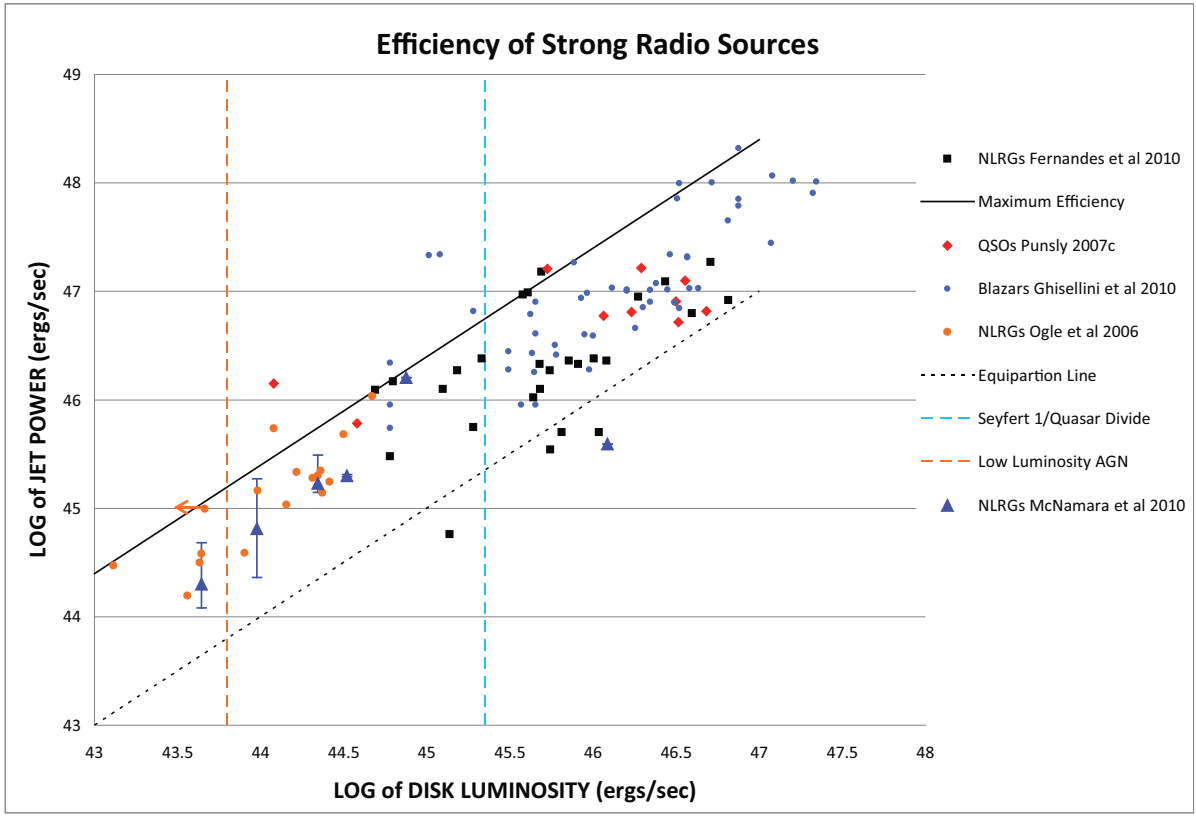}
\caption{The maximum efficiency line is illustrated in this scatter
plot of $Q$ and $L_{bol}$. The scatter plot shows that $1<R<25$
is not an unusual state of jet activity.}
\end{figure}
\par One can expand the \citet{fer10} treatment to a larger range of black hole accretion states and jet power,
by considering the low redshift sample of IR observations of FRII
NLRGs of \citet{ogl06}. The same IR bolometric correction and
$\overline{Q}$ estimators can be used as in \citet{fer10}. They noticed a diagonal boundary
in a scatter plot of IR luminosity versus 178 MHz flux which is similar to the maximum
efficiency line. In Figure 1, the orange circles represent these "weak
Mid-IR sources" from \citet{ogl06}. Notice how well the orange circles respect the
maximum efficiency line. The trend now extends below the upper limit
of LLAGN luminosity. Since Seyfert 1 galaxies also exist at such
luminosity, and the trend looks smooth, there does not seem to be
any evidence of a change in accretion mode for FRII NLRGs at
$L_{bol} < 6 \times 10^{43}$ ergs/sec.
\par A small sample of FRII NLRGs
from \citet{mcn10} is also plotted as blue triangles in Figure 1. In this
sample, $\overline{Q}$ is estimated by an independent method, the
work done to create large bubbles in the intra-cluster medium. The
IR luminosity is estimated from the data in \citet{shi05} with the
synchrotron component subtracted off and the IR bolometric
correction is from \citet{ric06}. This data also conforms with the
maximum efficiency line concept.
\par Another method of quantifying a jet as highly efficient is to choose sources with
$\overline{Q_{Edd}}=\overline{Q}/L_{Edd}>1$. This is an extreme
condition since $L_{bol}/L_{Edd}>1$ quasars are either extremely
rare or as is often argued nonexistent \citep{mar09,net09}. Thus,
the $\overline{Q_{Edd}}$ sources would almost certainly have
instantaneous episodes with $R(t)>1$. A small sample of these
sources were found in \citet{pun10}. These are the red diamonds in
Figure 1 and overlap the high end of the \citet{fer10} scatter.
They are lobe dominated quasars for which UV continuum emission and
broad line strengths were used to estimate $L_{bol}$ and 151 MHz
flux was used to estimate $\overline{Q}$. Note that the
\citet{fer10} and the \citet{pun10} samples are consistent with a
maximum value of $\overline{Q}\gtrsim 10^{47}$ ergs/sec that seems
to make the maximum efficiency line bend over towards the
equipartition line defined by $R=1$.
\par The fact that $\overline{Q}$ seems to reach a maximum does not
necessarily mean that there are not even larger instantaneous jet
powers. By fitting broadband blazar spectra, from the radio band to
gamma rays, \citet{ghi10} believe that they have a method to extract
the jet power within a few light years of the central black hole in
a blazar - almost contemporaneous on cosmic time scales. The model
is one of a highly relativistic magnetized plasmoid propagating in
the radiation environment of the quasar. The inverse Compton
emission is necessarily modeled simultaneously with an accretion
disk model of the "big blue bump" for each source. The method is a
bit controversial because of the large Doppler enhancement in blazar
jets and the uncertainty that it introduces in the intrinsic
luminosity. The appeal of this method is that it is completely
independent of the techniques used for the other data in Figure 1.
In order not to clutter Figure 1, only the $R(t)>2$ blazars are plotted.
These $Q(t)$ estimates (that are
plotted as dark blue circles) also respect the maximum efficiency
line with only two outliers.
\par None of the methods used to create the data sets in Figure 1 is a rigorous justification of the maximum efficiency
line in isolation. However, the agreement that is achieved by these
independent experiments are strong scientific evidence in support of
the notion of the maximum efficiency line found in \citet{fer10}
that is now extended to over 4 decades in $Q$. The fact that this
line extends over the full span of FR II radio power is an
indication that this is a fundamental property of jet production
that is independent of accretion power. Another important aspect of
Figure 1 is that powerful FR II radio sources are plentiful in the
range $1<R<25$. Furthermore, since jet power is not steady over the
lifetime of the QSO, many of the sources below the $R=1$ line likely
have episodes in the high efficiency range $1<R<25$, i.e. this is
not an aberrant or outlier state of jet activity.
\begin{figure}
\includegraphics[width=160 mm, angle= 0]{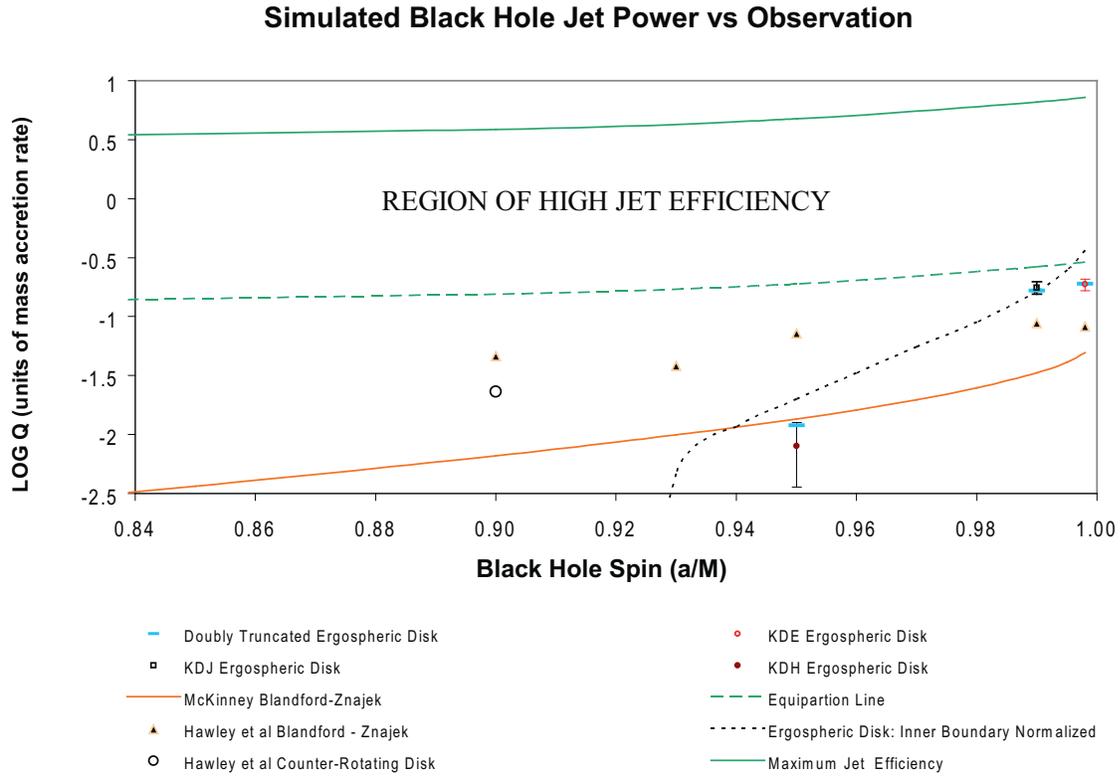}
\caption{A comparison of 3-D numerically simulated data with the
constraints on jet power in terms of accretion rate ($Q/\dot{M}
c^{2}$) imposed by observations. The "region of high jet efficiency"
corresponds to the region between $R=1$ and $R=25$ in Figure 1.}
\end{figure}

\section{Theoretical Discussion}
In the past, we were forced to compare a scatter plot like Figure 1
to theory by means of parametric models. However, these models have
a large unknown, the strength of the magnetic field
\citep{blz77,mei01,nem07}. The ability of a turbulent accretion disk
to transport and sustain large scale magnetic flux is controversial,
rendering the flux distribution as a major unknown
\citep{pra97,liv99,rot08,rey06}. Long term MHD simulations in a
generally relativistic background can at least provide a
self-consistent magnetic field distribution (be it not necessarily
unique). Thus, the best tools that we have to investigate the
central engine of radio loud AGN, without the over-riding
uncertainty of the field distribution, are the current battery of
long term 3-D numerical simulations of black hole accretion systems
\citep{mck09,bec08,haw06,kro05,fra07}. The initial state is a thick
torus of gas in equilibrium that is threaded by concentric loops of
weak magnetic flux that foliate the surfaces of constant pressure.
There is no net magnetic flux in these simulations. If the loops are
configured in the same orientation and are poloidal and not toroidal
then the leading edge that accretes will deposit a net poloidal flux
on the black hole and a jet forms \citep{bec08}. In these
simulations, $Q$ is expressible in terms of accretion rate onto the
black hole,  $\dot{M} c^{2}$. Thus, one can compare different
simulations and one can compare to the observations if $\eta_{th}$
is known. Theoretically, $\eta_{th}$ was determined by
\citet{nov73}. This was recently questioned since the boundary
condition of zero stress at the inner most stable orbit was suspect
\citep{gam99,kro99}. In spite of this, recent high resolution
simulations of accretion disks indicate that the \citet{nov73} value
is accurate to within a few percent for thin disks and only
increases modestly for high spin rates (expressed as $a/M \lesssim
1$, where "M" and "a" are the black hole mass and specific angular
momentum, respectively) and thick disks \citep{bec07,pen10,nob09}.
\footnote{Using the larger simulated $\eta_{th}$ values at high spin
rates will just elevate the equipartition and maximum jet efficiency
lines in the plot slightly, even farther from the already
nonconforming simulation data.} Therefore, in the context of this
discussion, the \cite{nov73} value of $\eta_{th}$ is suitable for
the purpose of comparing the simulations to the maximum efficiency
line in Figure 2. Recall that the sources in Figure 1 have $L_{bol}$
consistent with large viscous dissipation (big blue bump), so a
thermally luminous disk model is appropriate as opposed to low
efficiency advection dominated accretion \citep{nar94}.
\par Figure 2 compares the 3-D simulated data to the maximum efficiency line.
The \citet{mck09} a/M=0.92 simulation shows a
Blandford-Znajek (B-Z) jet as in \citet{blz77}. They note that $Q
\approx 0.01 \dot{M} c^{2}$ is similar to the 2-D solutions reported
in \citet{mck05}. Thus, the spin dependent efficiency equation,
equation (3) of \citet{mck05} is plotted above. The other B-Z jet
data (collectively referred to as Hawley et al. Blandford-Znajek in Figure 2)
comes from \citet{haw06,kro05} except for the a/M = 0.99 and the
a/M=0.998 data. The reason for the new data points is that the high
spin cases a/M=0.998 (KDE) form \citet{kro05,pun06} and a/M=0.99
(KDJ) from \citet{haw06,pun07,pun11} have a strong ergospheric disk
jet (as indicated in the Figure2) that suppresses the B-Z jet
\citep{pun08}. To find the power of a pure B-Z jet, the raw data
from two simulations (that were used for ray tracing in
\citet{bec07}) that was generously provided to this author by John
Hawley, KDEb (a/M=0.998) and KDJd (a/M=0.99), was reduced. These
simulations were different from KDE and KDJ because the code was
modified to include artificial diffusion terms in the equations of
continuity, energy conservation, and momentum conservation as
described in \citet{dev07}. The resultant numerical diffusion
suppressed the ergospheric disk jet giving a more pristine estimate
of the B-Z efficiency than can be obtained from KDE and KDJ. The simulation, KDH
(a/M=0.95), has a weak ergospheric disk jet so the estimates for the
B-Z power are straightforward \cite{pun08}. The
mechanical energy flux is included in these estimates which can be
non-negligible in some of the high spin simulations \citep{pun08}.
Note that no accretion disk jets form in any of these simulations.
\par Notice that all the no net flux simulations in Figure 2
fall below the region of high jet efficiency, $1<R<25$ from Figure
1, being 2 to 3 orders of magnitude less efficient than the most
efficient jets. The only simulations that are close are the high
spin ergospheric disk jets from KDE and KDJ. But they are still just
below the region of high jet efficiency. Can the a/M=0.998 case be
optimized to reproduce the high jet efficiency? To answer this, we
explore the curious situation that the ergospheric disk jet in KDE
is not much stronger than KDJ contrary to the theory \citep{pun01}.
A spin dependent expression for the ergospheric disk jet Poynting
flux (approximately $Q$ in this discussion), $S$, can be
approximated as
\begin{equation}
S \approx N(B^{2})\left[[SA]\Omega_{H}\right]^{2}\;,
\end{equation}
where the vertical magnetic field in the ergospheric disk is $B$,
the surface area of the ergospheric disk is $SA$ and the field line
angular velocity scales with the horizon angular velocity,
$\Omega_{H}$, and $N(B^{2})$ is a normalization constant
\citep{pun01}. The theory does not fix the distribution on $B$ in
time and space and $SA$ of the jet producing region that will occur
in a given numerical simulation. In equation (1), the empirical
distribution of $B^{2}$ is absorbed in the normalization constant
$N$. In principle, $SA$ of the equatorial plane in the ergosphere,
changes very rapidly with spin for $a/M \lesssim 1$, hence the
expectation of significantly larger jet power for KDE than for KDJ
\citep{pun01}. Setting $N= 0.0028$ in equation (1) creates the
"doubly truncated ergospheric disk," a two parameter ($SA$ and
$\Omega_{H}$) fit to the simulated data in Figure 2. It is doubly
truncated because the jet does not fill the entire ergospheric
equatorial plane, but is restricted to the region $r_{in}<r < 1.50
M$, where $r_{in}$ is the inner calculational boundary (located
outside the event horizon, $r_{in}>r_{H}$). For some unknown reason,
the vertical flux that creates the jet dies off rapidly beyond
$r=1.50 M$ in the simulations. Yet, in principle, an ergospheric
disk can exist throughout the ergosphere $r_{H} <r < 2 M$
\citep{pun01}. The simple two parameter equation (1) (with $N$
fixed) is a reasonable fit to the numerical data (the 3 blue dashes
in Figure 2) and explains the dependence of $S$ on $a$. Equation (1)
implies that the reason KDE is not much stronger than KDJ is because
$SA$ in the computational grids of the two simulations is virtually
identical ($\approx 18 M^{2}$ in geometrized units). If one were to
move $r_{in}$ in KDE inward so that it is normalized to the the KDJ
scaling of $r_{in}(KDJ)= 1.203 M = 1.054 r_{H}$, one expects a
larger $SA$ and therefore a larger jet power, (i.e. change the inner
calculational boundary in the numerical grid of KDE from
$r_{in}(KDE)=1.175 M$ to $r_{in} = 1.054 r_{H}= 1.12 M$). $SA$ is
increased to $23.3 M^{2}$ for a/M=0.998 if $r_{in}=1.12M <r < 1.5
M$. The black dashed curve in Figure 2 is a plot of the predicted
$S$ from equation (1) with the empirically fit value of $N = 0.0028$
from the simulated data and normalized numerical grids defined by
$r_{in} = 1.054 r_{H}$. Even with the enhancement in power for a/M =
0.998, with a renormalized grid, the curve barely penetrates the
region of high jet efficiency. Based on this analysis of this most
efficiently known simulated jet, it is concluded that any attempt to
optimize the no net flux scenario will still be too weak to explain
the region of high jet efficiency in Figure 2.
\section{Conclusion} This study shows that the maximum jet efficiency
condition, $R\approx 25$, extends over the entire range of known FR
II jet powers. It is demonstrated that the no net flux accretion
numerical models can not explain the large number of high efficiency
jets with $R>1$. Furthermore, the initial no net flux state in the
considered simulations is configured to yield the maximum jet power.
The situation for no net flux is even more nonconforming if the
initial field is not composed of loops of the same vertical
orientation, but randomly oriented. In this case, the jet power can
be reduced by three orders of magnitude or more \citep{bec08}. It is
concluded that the no net flux models are not suitable numerical
models for powerful radio loud quasars. Clearly some other dynamical
element is needed in the setup of the simulations. Perhaps it is the
accretion of a large reservoir of large scale magnetic flux as in
\citet{igu08}. This can have two relevant effects. First, the
ergospheric disk power should be greatly increased \citep{pun11}.
Also, unlike the no net flux simulations, a magnetized accretion
disk forms in \citet{igu08}.
\par Finally, note that Figure 2 does not indicate that large "a" acts likes a
switch that transforms a radio quiet black hole accretion system
into one with high jet efficiency. The only thing that looks
remotely like a switch in Figure 2 is the ergospheric disk. This
conclusion  is in accord with the B-Z switch model of \citet{tch10}
that requires radio quiet (loud) quasars to have spins $a \approx
0.15$ ($a \approx 1$). This scenario seems unlikely, radio quiet
quasars are typically selected by optical/UV luminosity and
therefore should have elevated (mass and angular momentum) accretion
rates and large $a$ \citep{bar70}. The magnetospheres in
\citet{tch10} are shaped by a boundary condition. There is no
accretion in the simulation, therefore there is no calibrated
measure of the jet strength with respect to $\dot{M} c^{2}$.
\begin{acknowledgements}
I would like to thank an anonymous referee who offered many
constructive comments.
\end{acknowledgements}

\end{document}